\documentstyle[12pt,epsf,russian]{article}
\newcommand{\ds}{\displaystyle}
\newcommand{\dsf}{\ds\frac}

\newcommand{\beq}{\begin{equation}}
\newcommand{\eeq}{\end{equation}}
\newcommand{\onehalfspace}
{\renewcommand{\baselinestretch}{1.}
\large\normalsize}
\onehalfspace
\begin{document}
\begin{center}
{\Large\bf
THERMOMAGNETIC SHOCK WAVES IN THE VORTEX STATE OF TYPE-II SUPERCONDUCTORS}
\vskip 0.1cm
{\normalsize\bf N.A.\,Taylanov}\\
\vskip 0.1cm
{\em Theoretical Physics Department,\\
Institute of Applied Physics,\\
National University of Uzbekistan,\\
E-mail: taylanov@iaph.tkt.uz}
\end{center}
\begin{center}
{\bf Abstract}
\end{center}
\begin{center}
\mbox{\parbox{14cm}{\small
       The nonlinear dynamics of thermal and electromagnetic perturbations
in the vortex state of type II superconductors is analyzed with account
of dissipation and dispersion effects.
A theoretical analysis shows that nonlinear thermal and electromagnetic
dissipative waves (structures) may be formed under certain conditions on the
sample surface. The structures possess a finite-amplitude
and propagate at a constant velocity. The appearance of these structures
is qualitatively described and the wave propagation velocity is estimated.
The possibility of experimental observation of thermomagnetic shock waves is
briefly discussed.
}}
\end{center}
\vskip 0.5cm
\begin{center}
{\bf Introduction}
\end{center}
\vskip 0.5cm

          Recent years a greater number of comparatively examples
of spontaneous origin of spatial and temporary dissipative waves
(structures) in unordered systems - self-organization processes  -
become well known.
By the terminology of Prigozhin [1] dissipative structures are organized
condition in the space and time, which can move over to the thermodynamic
balance condition of only by jumping (as a result of kinetic phase
transition). If the deflection of a nonlinear system from the balance
exceeds a critical value, these conditions can become unstable. In this case
the system moves over to the new mode and becomes a dissipative structure,
which  appears and exists due to the dissipative  processes. There are
different types of dissipative structures [2, 3].

        At present, superconducting systems with high critical field
strengths and current densities are widely implemented in various advanced
technologies. However, successful operation of the superconducting
materials is only possible provided that special measures are taken
to prevent a system from the thermal or magnetic breakage of
superconductivity and the transition to a normal state.
For this reason, one of the main problems in the investigation of
properties of superconductors is predicting of the superconducting
state breakage caused by dissipative and nonlinear effects related to
viscous motions of the magnetic flux. This explains the considerable
interest in the study of dissipative and nonlinear effects in
superconductors that has arisen in recent years.

        The existence of essential nonlinear and dissipative effects,
connected with Joule heating at the moving of magnetic flow, creates
different regimes of automodel processes, describing the evolution of
thermal and electromagnetic perturbations in the vortex (resistive)
state of type-II superconductor. One of measured regimes is the propagation
of thermomagnetic waves (normal region - the region heated with the
temperature, which is higher than the critical temperature $T_c$) [3].
In early investigations [3, 4] it was shown that the nonlinear stage
of evolution of thermal and electromagnetic perturbations in superconductors
is determined by the stationary profile in the form of a running wave.

       In the present work the nonlinear dynamics of thermal and
electromagnetic perturbations evolution in the vortex state of
type II superconductors is investigated with the account of dissipation
and dispersion effects. It was found the existence of nonlinear
thermomagnetic waves, describing the final stage of evolution of thermal and
electromagnetic perturbations in the vortex state of superconductors.
An estimate for the width and the velocity of a propagating wave is obtained.
\vskip 0.5cm
\begin{center}
{\bf \S 1. Running wave in the vortex state of superconductors}
\end{center}
\vskip 0.5cm

         The evolution of the thermal $(T)$ and electromagnetic
$(\vec E, \vec H)$ perturbations in the vortex state of
superconductors is described by the nonlinear equation
of the thermal conductivity [4]
\beq
\nu\frac{dT}{dt}=\nabla [\kappa\nabla T]+\vec j\vec E,
\eeq
by the Maxwell equations
\beq
rot\vec E=-\dsf{1}{c}\dsf{d\vec H}{dt},
\eeq
\beq
rot{\vec H}=\dsf{4\pi}{c}\vec j
\eeq
and by the equation of the vortex state
\beq
\vec j=\vec j_{c}(T,\vec H)+\vec j_{r}(\vec E),
\eeq
here $\nu=\nu(T)$ and $\kappa=\kappa(T)$ are the heat capacity and the
thermal conductivity respectively; $\vec j_c =\vec j_c (T,\vec H)$ is the
critical current density and $\vec j_r=\vec j_r(\vec E)$ is the resistive
current density.

        The above system is essentially nonlinear because the right-hand
part of Eq.(1) contains a term describing the Joule heating
in the region of the resistive phase.
Such a set (1)-(4) of nonlinear parabolic differential equations in
partial derivatives has no exact analytical solution.

        Let us consider a planar semi-infinite sample $(x>0)$ placed in
external magnetic field $\vec H=(0, 0, H_{e})$ growing at a constant rate
$\dsf{d\vec H}{dt}$=const. According to the Maxwell equation (2), there
is a vortex electric field $\vec E=(0, E_e, 0)$ in the sample, directed
parallel to the current density $\vec j$:
$\vec E\parallel \vec j$; here $H_e$ is the amplitude of the external
magnetic field and $E_e$ is the amplitude of the external electric field.

        In this section we will discuss the question on possibility
of existence of automodel solution to the initial system of equations (1) -
(4) in private derivatives with independent variables $x$ and $t$ in the
form of running impulse, describing the nonlinear evolution of thermal and
electromagnetic perturbations in superconductors.

        The system of equations, describing the dynamics of evolution of
thermomagnetic perturbations in the vortex state of superconductors
is essentially nonlinear and it is impossible to integrate them in this
form. As well known that, it is difficult to solve
the problem with initial and boundary conditions for nonlinear differential
equations in private derivatives by common methods, so that we can not use
the superposition principle.
Therefore it is reasonable to search other ways of the solution.
There is such method, which allows  the transformation of resemblance, due to this the
system of equations with private derivatives can be reduced to the system
of common differential equations. Recent years more attention is given to
investigation of such automodel solution [5]. The obtained by the method
differential equations with common derivatives can be integrated in the
close form or reduced to investigation of one equation, which can be solved
either analytically or numerically.

In many problems of mathematical physics the automodel solution of running
wave type is of great interest [5]. For important class of stationary
running waves the distribution of characteristics of waves is still
unchangeable on time. In accordance with this determination the solution
in the form of running wave can be presented in the form
\beq
u=u(x\pm vt),
\eeq
where $u$ - the characteristics of the event under consideration, $x$ -
coordinate, $t$ -
time, $v$ - the velocity of the running wave. The solution (5) presents
a wave, running either in negative or positive direction of $x$ with
constant velocity $v$.

The possibility of reducing of the system of differential equations in
private derivatives to the system of common differential equations in
automodel derivatives greatly simplifies the problem in the mathematical
point of view and in many cases allows to find exact analytical solution.
Let us consider the problem about in what conditions the system of initial
differential equations in private derivatives with independent variables
$x$ and $t$ can be reduced to the system of differential equations in common
derivatives, depending on only one variable $\xi(x,t)$.
The solution of the system of equations (1)-(4) may be presented as a
function of new automodel variable $\xi(x,t)$:
\beq
\begin{array}{l}
T=T[\xi(x,t)],\\
\quad\\
E=E[\xi(x,t)],\\
\quad\\
j=j[\xi(x,t)].\\
\end{array}
\eeq

        Substituting (6) into the system (1)-(4) gives, as a result of simple
differentiation, the following system

\beq
\dsf{d\xi}{dt}\left[\nu\dsf{dT}{d\xi}\right]=
\kappa\left\{\dsf{d^2\xi}{dx^2}\dsf{dT}{d\xi}+
\left(\dsf{d\xi}{dx}\right)^2
\dsf{d^2T}{d\xi^2}\right\}+[j_c(T)+j_r(E)]E\,,
\eeq

\beq
\dsf{d^2\xi}{dx^2}\dsf{dE}{d\xi}+\left(\dsf{d\xi}{dx}\right)^2
\dsf{d^2E}{d\xi^2}=\dsf{4\pi}{c^2}
\left[\dsf{dj_c}{dT}\dsf{dT}{d\xi}+\dsf{dj_r}{dE}\dsf{dE}{d\xi}\right]
\dsf{d\xi}{dt}\,.
\eeq

       In order the system (7),(8) was only function from $\xi$
 at the substation (6) it is required carrying out the following conditions:
\beq
\dsf{d\xi}{dt}=A(\xi)\,,
\eeq
\beq
\left(\dsf{d\xi}{dx}\right)^2=B(\xi)\dsf{d\xi}{dt}=G(\xi)\,,
\eeq
\beq
\dsf{d^2\xi}{dx^2}=C(\xi)\dsf{d\xi}{dt}\,,
\eeq
where $A,C,G$ are functions to be determined below. Solving first two
system of equations (9)-(11) we have a relation
\beq
G(\xi)\dsf{dA}{d\xi}=A(\xi)\dsf{dG}{d\xi}\,.
\eeq
Hence just follows relationship between $G$ and $A$
\beq
G(\xi)=\dsf{1}{u}A(\xi)\,,
\eeq
here $u$ is a free constant of integrating of the equation (12). From
(9) and (10) follows that $\xi(x,t)$ must satisfy single-line equation
in private derivation
\beq
\dsf{d\xi}{dt}=u\dsf{d\xi}{dx}
\eeq
the only solution of which is the function
\beq
\xi(x,t)=F(x-ut)\,.
\eeq

        Using (15) we can immediately obtain

\beq
G(\xi)=1,\quad A(\xi)=-Fu,\quad C(\xi)=0.
\eeq
It is possible to ensure $F=1$ by the transformation of coordinates and times.
Thereby, we find final automodel substitution
\beq
\xi=x-ut,
\eeq
corresponding to solution of a running type wave [5].
\vskip 0.5cm
\begin{center}
{\bf \S 2. Differential equations for nonlinear $E$ and $H$ waves}.
\end{center}
\vskip 0.5cm

        For the automodeling solution of the form (17), describing
a running wave moving at a constant velocity $v$ along the $x$ axis,
the system of equations (1)-(4) takes the following form
\beq
-\nu (T) v\dsf{dT}{d\xi}=\dsf{d}{d\xi}\left[\kappa (T)\dsf{dT}{d\xi}\right]
+[j_c(T)+j_r(E)]E,
\eeq
\beq
\dsf{d^2E}{d\xi^2}=-\dsf{4\pi v}{c^2}\dsf{dj}{d\xi},
\eeq
\beq
\dsf{dE}{d\xi}=\dsf{v}{c}\dsf{dH}{d\xi}.
\eeq

       The thermal and electrodynamics boundary conditions for equations
(18)-(20) are as follows:
\beq
\begin{array}{l}
T(\xi\rightarrow+\infty)=T_0,\quad \dsf{dT}{d\xi}(\xi\rightarrow-\infty)=0,\\
\quad\\
E(\xi\rightarrow+\infty)=0,  \quad E(\xi\rightarrow-\infty)=E_e,\\
\end{array}
\eeq
here $T_0$ is the temperature of the cooling medium.

        In this section in view of considerable analytical difficulties
involved in solving the exact problem, we will restrict our consideration to
the so-called Bean-London critical state model [6] by assuming that
$j_á=j_c(T, H_e)=j_c(T)$ and
\beq
j_{c}(T)=j_0-a(T-T_{0}),
\eeq
where $j_{0}$ is the equilibrium current density,
$a$ is the thermal heat softening coefficient of the magnetic flux pinning
force [7].

        The characteristic field dependence of $j_r(E)$ in the region of
sufficiently strong electric field $(E\ge E_f)$ can be approximated by a
piece-wise linear function $j_r\approx\sigma_f E$, where
$\sigma_f=\dsf{\eta c^2}{H\Phi_0}\approx \sigma_n \dsf{H_{c_2}}{H}$ is the
effective conductivity in the flux-flow regime; $\eta$ is the viscous
coefficient,\quad $\Phi_0=\dsf{\pi h c}{2e}$ is the magnetic
flux quantum, $\sigma_n$ is the conductivity in the normal state;
$H_{c_2}$ is the upper critical magnetic field,\quad
$E_f$ is the boundary of the linear area in the voltage-current
characteristics of the sample [8].
In the region of small field strengths $(E\le E_f)$, we assume a relationship
$j_r(E)\approx j_1\ln\dsf{E}{E_0}$ to be valid with $j_1$ being a
characteristic local current density scatter (related to a pinning force
inhomogeneity [9]) on the order of $j_1\approx 0,01j_c$; $E_0$=const.
This relationship between $j$ and $E$ is due to a thermoactivated flux-creep.
The thermoactivated flux-creep is a principal mechanism of the energy
dissipation during the magnetic flux penetrates deeply into the sample.

     Excluding variables $T(\xi)$ and $H(\xi)$ from Eqs. (18) and (20) ,
and taking into account the boundary conditions (21), we obtain an equation
describing the electric field $E(\xi)$ distribution ($E$-wave):
\beq
\dsf{d^2 E}{d\xi^2}+\left[\dsf{4\pi v}{c^2}\dsf{dj_r}{dE}\dsf{dE}{d\xi}+
\dsf{4\pi v^2a}{c^2}\dsf{N(T)-N(T_0)}{\kappa (T)}\right]-
\dsf{aE^2}{2\kappa (T)}=0,
\eeq
where the dependency $T=T\left(E,\dsf{dE}{d\xi}\right)$ is defined by
expression (18), (19) and have the form
\beq
T=T\left(E,\dsf{dE}{d\xi}\right)=T_0+\dsf{1}{a}\left[j_0+j_r(E)+
\dsf{4\pi v}{c^2}\dsf{dE}{d\xi}\right].
\eeq
Here $N(T)=\int\limits_{0}^{T}\nu(T)dT$.
The velocity of  $E$-wave is determined by the Eq. (23) with account
of the boundary conditions
\beq
v_{E}^{2}=\dsf{c^2}{8\pi}\dsf{E_{e}^{2}}
{N\left[T_0+\dsf{1}{a}[j_c(T)+j_r(E)]\right]-N(T_0)}.
\eeq
Using Eqs. (20) and (23) one can easily find the expression for
distribution of the magnetic field $H$ in the case of $H$-wave
\beq
\dsf{d^2H}{d\xi^2}+\dsf{4\pi v}{c^2}\left[\left.\dsf{dj_r}{dE}
\right|_{E=\dsf{v}{c}H}\dsf{dH}{d\xi}+ca\dsf{N(T)-N(T_0)}
{\kappa (T)}\right]-\dsf{av}{2c}\dsf{H^2}
{\kappa (T)}=0.
\eeq
The velocity $v_H$ of $H$-wave is connected with its amplitude $H_e$ by the
following expression
\beq
N(T)-N(T_0)=\dsf{H_{e}^{2}}{8\pi}.
\eeq
   The expression (27) shows the adiabatic nature of the $H$ - wave
propagation: the magnetic energy, accumulated during decay of the
field, is spent for heating the superconductor in the region close to
the wave front. Therefore, depending on the character of the external
electrodynamics conditions on the surface of the sample there may exist
two type (E and H) of nonlinear stationary waves in the superconductor.
\vskip 0.5cm
\begin{center}
{\bf \S 3. Thermomagnetic shock waves}
\end{center}
\vskip 0.5cm

      The first integrals of equations (18)-(20) have the form

\beq
-\nu v(T-T_0)=\kappa\dsf{dT}{d\xi}-\dsf{c^2}{8\pi}E^2\,,
\eeq
\beq
j=-\dsf{c^2}{4\pi v}\dsf{dE}{d\xi}\,.
\eeq
\beq
E=\dsf{v}{c} H\,,
\eeq
(an integrating constants in Eqs. (28)-(30) are equal
to zero on the strength of the boundary conditions at $\xi=-\infty$).
Using integrals (28), (29) and excluding variable $T(\xi)$ we can get
the following differential equation for the distribution of $E(\xi)$:
\beq
\dsf{d^2 E}{dz^2}+\beta(1+\tau)\dsf{dE}{dz}+\beta^2\tau E
-\dsf{E^2}{2E_\kappa}=0\,.
\eeq
Here the following dimensionless parameters are introduced:
$z=\dsf{\xi}{L},\quad \beta=\dsf{vt_\kappa}{L}$, \quad
$L=\dsf{cH_e}{4\pi j_0}$ is the depth of the magnetic field penetration
into the superconductor, $\tau=\dsf{D_t}{D_m}$ is the parameter
describing the ratio of the thermal $D_{t}=\dsf{\kappa}{\nu}$ to the magnetic
$D_{m}=\dsf{c^2}{4\pi\sigma_{d}}$ diffusion coefficient,
$t_\kappa=\dsf{\nu L^2}{\kappa}$ is the time of thermal diffusion, and
$E_{\kappa}=\dsf{\kappa}{aL^2}$ is a constant parameter.

        One can see that the mathematical model described by Eq. (31)
can be considered as a decreasing nonlinear oscillator at the presence of a
friction force
\beq
F_t=-\beta(1+\tau)\dsf{dE}{dz}\,,
\eeq
(where $z$ is the analogue of time, $E$ is the coordinate of a "material
point") with a potential [5]
\beq
U(E)=-\dsf{E^3}{6E_\kappa}+\beta^2\tau\dsf{E^2}{2}.
\eeq
    The analysis of the phase plane of Eq. (31) shows that there are two
equilibrium points in it: $E_0$ is a stable node and $E_1$ is a saddle
node. These two equilibrium points  $(E_0,E_1)$ are separated in the phase
plane by the separatrix $AB$ (Fig.1). At $z\rightarrow +\infty$ the
material point is in the point $E_1$, and at  $z\rightarrow -\infty$
it moves to the point $E_0$. The transition from one equilibrium state
to another one occurs only monotonycally. This solution by joining these
two equilibrium points may be presented in the form of the shock-wave-type
with amplitude $E_e$ running at a constant velocity $v_E$ (Fig.2).

Let us consider the solution of the Eq (31), corresponding to
the limiting case $\tau<<1$ (hard superconductors).
It follows from the $\tau<<1$ condition that the thermomagnetic perturbation
develops nearly adiabatically [8]. Therefore in this approximation the
thermal conductivity is negligible and corresponding terms, containing
dissipative effects can be neglected in Eqs. (28)-(30).
Then for the linear region of the current-voltage characteristic
the solution to Eq. (31) can be presented in the form
\beq
E(z)=\dsf{E_1}{2}\left[1-\rm{tanh}\dsf{\beta\tau}{2(1+\tau)} (z-z_0)\right]\,,
\eeq
Here $z_0$ is a constant.
The condition $\tau<<1$ means that the magnetic flux diffusion $D_t$
is considerable faster than that of the thermal diffusion $D_m$.
Then we can assume that the spatial scale of the variation of the magnetic
field penetration $L_E$ is essentially larger than the
corresponding thermal scale $L_T$, therefore the spatial derivatives
$\dsf{d^n E}{dz^n}$ contain the small parameter
$\dsf{L_T}{L_E}<<1$.
It is easy to check fairness of this approach by differentiating Eq.
(31):
\beq
{\dsf{d^2 E}{dz^2}}\left(\beta{\dsf{dE}{dz}}\right)^{-1}=\tau<<1.
\eeq
Maximum inaccuracy of given approximations is of the order of
$\dsf{\tau}{(1+\tau)^2}$.
For example, for  $\tau=1$ it is
25 \%, and in the limiting case $\tau\rightarrow 0$  or $\tau\rightarrow
\infty$ is very small.

        The expression (34) describes the structure of the thermomagnetic
shock wave, penetrating into the superconductor (see, Fig.2).
      Using the boundary
conditions for $E$-wave one can easily find the wave velocity
\beq
v_E=\dsf{L}{t_\kappa}\left[\dsf{E_e}{2\tau E_\kappa}\right]^{1/2},
\eeq
with amplitude $E_e$ and with front width
\beq
\delta z=16\dsf{1+\tau}{\tau^{1/2}}\left[\dsf{E_\kappa}{E_e}\right]^{1/2}.
\eeq

         Numerical estimations give the value
$v_E=10^1\div 10^3 \cdot\dsf{\rm{sm}}{\rm{sec}}$ and $\delta z=10^{-1}\div 10^{-2}$ for
$\tau=1$.

        Thus, the nonlinear stage of evolution of thermal and electromagnetic
perturbations in the vortex state of superconductors is determined by the
stationary profile in the form of running wave. It seen that, the transition
from superconducting state to normal can occur by the expanding of stationary
thermomagnetic wave, which is caused by the balance between nonlinear
and dissipative effects, caused  by Joule heating at the viscous motion
of Abrikosov vortices inside superconductor.
\vskip 0.5cm
\begin{center}
{\bf \S 4. Nonlinear waves in the flux-creep regime of hard superconductors}
\end{center}

        As mentioned in Sec.2 in the viscous flux-flow regime
the electric conductivity $\sigma_f$ is independent of the
electric field $E$, the profile of thermomagnetic waves is independent
of the background electric field in the sample.
In the region of small electric field strengths the
current-voltage characteristic of the sample has a nonlinear part and
it is determined by the following logarithmic dependence
$$
j_r(E)\approx j_1\ln\dsf{E}{E_0}.
$$
The current-voltage characteristics of hard superconductor is described
by the last equation over a wide interval of temperatures $T$ and magnetic
fields $H$.
This factor may considerably affect the character of nonlinear
thermomagnetic wave propagation in the sample.
In this section we consider a problem of thermomagnetic wave profile,
with account the nonlinear current-voltage curve of superconductor.
Taking into account the nonlinear dependence $j_r(E)$, we can represented
Eq. (24) in the following form
$$
T=T_0+\dsf{1}{a}\left[j_0+j_1\rm{ln}\dsf{E}{E_0}+\dsf{4\pi v}{c^2}
\dsf{dE}{dz}\right].
$$
Substituting the last expression into Eq. (18), and taking into account
the boundary conditions (21), we obtain an equation describing the
distribution of $E$ - wave:
\beq
\dsf{d^2E}{dz^2}+\beta\left[1+\dsf{j_1}{\sigma_d(E) E}\tau
\right]\dsf{dE}{dz}+\beta^2\tau
\left[\dsf{j_0}{\sigma_d(E)}+\dsf{j_1}{\sigma_d(E)}\ln\dsf{E}{E_0}
\right]=\dsf{E^2}{2E_\kappa}\,.
\eeq
Here $\sigma_d=\sigma_d(E)$ is the differential conductivity in the
flux-creep regime [8].
The corresponding equation of state is obtained using the relationship [10]
\beq
\Omega E^2=X(E)=1+\dsf{j_1}{j_0}\ln\dsf{E}{E_0}\,,
\eeq
where
$\Omega=2j_0\sigma_d^{-1}(E_e)E_\kappa\beta^2\tau$ is a constant.
As seen from the plots of $X(E)$ versus $E$ (Fig.3), there exists
a single intersection point of the curves $y=\Omega E^2$ and $y=1+\dsf{j_1}{j_0}\ln\dsf{E}{E_0}$, which
corresponds to a single stable equilibrium state. The stability of this
state is determined by the sign of the derivative $\dsf{d^2E}{dz^2}$ in the
vicinity of the equilibrium point. The
wave velocity $v_E$ can be determined from Eq. (39) using for the
boundary conditions and have the form of
\beq
v_E=\dsf{L}{t_\kappa} E_e\left[2\tau\dsf{j_0E_\kappa}{\sigma_d(E)}\left(1+
\dsf{j_1}{j_0}\ln\dsf{E_e}{E_0}\right)\right]^{-1/2}\,.
\eeq
Now we can use Eq. (30) and readily derive an equation describing the
field distribution for a nonlinear $H$-wave. The wave velocity $v_H$ of the
$H$-wave is given by the formula
\beq
v_H=\dsf{cE}{H_e}
\exp\left[
\dsf{\sigma_d(E)}{2\tau j_1}E_\kappa\left(\dsf{LH_e}{ct_\kappa}\right)^2
-\dsf{j_0}{j_1}\right]\,.
\eeq
It is seen that the wave velocity increases exponentially with the amplitude
$H_e$. For a sufficiently small amplitude $H_e<H_a$, where
\beq
H_a=\dsf{ct_\kappa}{L}\left[\dsf{2\tau j_0E_\kappa}{\sigma_d(E)}\right]^{1/2}
\eeq
the wave velocity is negligible small, which corresponds to the case
of the thermoactivated flux creep. For $H_e=H_a$ the wave propagates at a
finite constant velocity $v_H$.
The expression
\beq
E=E_0\exp\left[\dsf{\sigma_d(E)}{2\tau j_1}E_\kappa\left(\dsf{LH_e}
{ct_\kappa}\right)^2-\dsf{j_0}{j_1}\right]\,.
\eeq
represents the intensity of the spontaneous electrical field caused by the
thermally activated magnetic flux-creep inside the region warmed up during
wave propagation.
In this case the maximum heating of the superconductor in the region
immediately close to the wave front is described by the relationship
$$
\dsf{T-T_0}{T_0}=\Theta=\dsf{H_{e}^{2}}{8\pi \nu T_0}.
$$
Let us now estimate the maximum heating $\Theta$
from the following typical values of the physical parameters of the sample;
$H_e=10^3\cdot \rm{Gs}$, $\nu=10^5\cdot\rm{erg}/(\rm{cm^3}\cdot \rm{K})$,
$T_0=4.5 \cdot\rm{K}$, we have $\Theta\approx 0.2\sim 0.5 $. At this value of
the heating the wave velocity is of the order of $v_H\approx 10\sim 100\cdot
\rm{sm}/\rm{sek}$.
Finally it should be note that taking into account the nonlinear dependence
of the current density $j$ on the electric field strength $E$ should not
qualitatively change the main results, since the character of the
equilibrium state on the phase plane remains essentially the same.
\vskip 0.5cm
\begin{center}
{\bf \S 5. The structure of waves}
\end{center}
\vskip 0.5cm

        The nature of the structure of the nonlinear $E$ (or $H$) - wave
can be determined by investigating the asymptotical behaviour
at $z\to \pm\infty$ of solutions for $E(z)$ in the
close vicinity of the equilibrium points $E=E_e$. Linearizing the
equation (38) near the equilibrium point (weak-nonlinear wave) it can
be represented in the following form
 \beq
\dsf{d^2 E}{dz^2}+\beta(1+\tau)\dsf{dE}{dz}+
\left[1-\dsf{\sigma_d(E_e)E_e}{2j_0(T_0,E_e)}\right]\dsf{(E-E_e)}
{2E_\kappa}=0.
\eeq
The analysis of the phase plane ($E,\dsf{dE}{dz}$) of Eq.(44) shows that
there is a single equilibrium point, which is a saddle under the condition
$$
j_0(T_0,E_e)>\dsf{1}{2}\sigma_d(E_e)E_e.
$$
        The spatial scale of the variation of the solutions of Eq.(44),
obviously, determines the width of the wave front $\Delta z_0$:
\beq
\delta z_0=\left[1-\dsf{\sigma_d(E_e)E_e}{2j_0(T_0,E_e)}\right]^{-1}
[(1+\tau)\beta]^{-1}.
\eeq
An asymptotical solution of the Eq. (44) can be found in the field of
$z\to +\infty (E\to 0)$ using the voltage-current characteristics
\beq
j\to E^{1/n}; \quad (n>>1),
\eeq
in the flux-creep regime $E<<E_0$ of the following form
\beq
E(z)=E_1 \exp(-n\beta z).
\eeq
The total width of the wave front is determined obviously, by the value
$\Delta z_0$ since in the region of $z>0$ (on the strength of the condition
$n>>1$) the amplitude of the wave sharply tends to zero
(on the scale $\Delta z^*\approx\dsf{1}{n\beta}<<\Delta z_0)$.
It is seen that the width of the $E$-wave front is determined mainly by the
largest of the parameter heat capacity $\nu$ and thermal conductivity
$\kappa$ coefficients ($\beta\sim\dsf{\kappa}{\nu}$).
\vskip 0.5cm
\begin{center}
{\bf \S 6. Waves with the finite amplitude}
\end{center}
\vskip 0.5cm

        It is notice able that the evolution of perturbations of
temperature $T(x,t)$ and fields $E(x,t)$ and $H(x,t)$ essentially
is determined by the equation of resistive state (4) and physical
parameters of the sample.
It should be noted that inclusion of the temperature dependences of the
parameters $\kappa$ and $\nu$ substantially complicates analytical
calculations of the wave propagation dynamics that is described by the
system of Eqs. (28)- (30). In most cases, the changes in the local
values of these parameters in the sample can be considered negligible
comparatively to the characteristic scale of temperature variations.
Hence, we can take these parameters to be constant.
Indeed, the investigation revealed that the thermal conductivity almost
does not affect the character of the stationary
wave propagation. This stems from the fact that the thermal flux
$\kappa\dsf{dT}{dz}$ vanishes at stationary points of the system
at $z\rightarrow\pm \infty$. However, the
temperature dependence of the heat capacity should be taken into account.
Such a dependence is represented as
$\nu\approx\nu_0\left[\dsf{T}{T_0}\right]^{3}$ over a wide range of
temperatures [8].
Then Eq. (28) can be written in the following form
$$
-\nu_0 v\left[\dsf{T-T_0}{4T_0}\right]^4=\kappa\dsf{dT}{dz}-\dsf{c^2}{8\pi v}E^2.
$$
By eliminating variable $T(z)$ from the last relationship and employing
the boundary conditions (21), we obtain
\beq
\dsf{d^2 E}{dz^2}-2\pi\dsf{\nu T_0}{E_\kappa}\dsf{v^2}{c^2}\left[\left(1+
\dsf{\sigma_f E}{aT_0}+\dsf{c^2}{4\pi avT_0}\dsf{dE}{dz}\right)^{4}-1
\right]+
\beta\tau\dsf{dE}{dz}=\dsf{E^2}{2E_\kappa}\,.
\eeq
        According to the qualitative theory [11], the equilibrium states are
found from the condition
\beq
2\pi\nu_0 T_0\dsf{v^2}{c^2}
\left[\left(1+\dsf{\sigma_f E}{aT_0}\right)^{4}-1\right]=E^2\,.
\eeq
An evident property of system (49) is the absence of closed curves that
are fully composed of the phase trajectories in the phase plane
$\left(E,\dsf{dE}{dz}=P\right)$.
The proof of this statement can be based on the Bendixson criterion [10].
The number of stationary points (one or three) and their
type are determined by the parameter
\beq
W=2\pi\nu_0T_0\dsf{v^2}{c^2}\,.
\eeq
The three equilibrium points $E=0$, $E=E_1$ and $E=E_2$ correspond to the
condition $W<W_k=\dsf{1}{2}$ (Fig.4). There is only one singular point
$E_0=0$ at $W>W_k$. The parabola and the quadratic curve in Eq. (49) are
tangent at $W=W_k$; i.e., this condition corresponds to the coincidence
$E_1=E_2=E^*=\dsf{6}{7}\dsf{aT_0}{\sigma_{f}}$.
The direct solution of Eq. (49) yields the following waves:
\begin{eqnarray}
\begin{array}{lll}
1)& E_{1,2}=E^*[1+2,2(W_k-W)^{1/2}]\,,&\quad\mbox{at}\quad
\left(\dsf{W_k-W}{W_k}\right)<<1\,;\\
\quad\\
2)& E_1=8\pi\dsf{\sigma_f\nu_0}{a}\dsf{v^2}{c^2}\,,&\\
\quad\\
  & E_2=(2\pi)^{1/2} \dsf{\sigma_f^2\nu_0^2}{a^2}\dsf{c}{v}>>E_1\,,&
\quad\mbox{at}\quad W_k>>W\,.
\end{array}
\end{eqnarray}
       Analysis of the phase plane $(E, P)$ shows
that the points $E_0=0$ and $E=E_2$
are stable nodes and that $E=E_1$ is a saddle. In addition to the separatrix
$E_1 E_0$, the set (49) has the separatrix $E_1 E_2$ connecting the points
$E_1$ and $E_2E_1$. This means that two types of waves with
amplitudes
$\Delta E=E_1$ and $\Delta E=E_2-E_1$ can exist in the superconductor (Fig.5).
Evidently, wave I has an amplitude of the order $E_k$ at $W\rightarrow W_k$;
its velocity is determined by equality (49) at
$E=E_{1}$. Equation (49) has two stationary points at $W<<W_k$:  $E_{0}=0$
is a stable node and $E_{1}=2\beta^2\tau E_{\kappa}$ is a saddle.
The separatrix that connects these two equilibrium states corresponds to a
"difference"-type solution with amplitude $E_{e}$, which is related to
the wave
velocity  $v_{E}$ and the wave front width $\Delta z$ by the equations
(36) and (37), respectively.
Wave II has a small amplitude at $W\rightarrow W_k$
\beq
\dsf{\Delta E}{E_k}=4,4(W_k-W)^{1/2} <<1\,,
\eeq
and its velocity is inversely proportional to the amplitude at $W<<W_k$.
Such an exotic dependence of $v_E$ on $\Delta E =E_e$ most likely means that
the waves of this type are unstable.
Finally it is notice able that observation of the
second-type waves becomes possible in finite-sized samples with asymmetric
boundary conditions. The above investigations prove the possibility of
applying the results obtained to high-temperature
superconductors cooled to liquid-nitrogen temperatures (T = 77 K), providing
that the values of the physical parameters of the sample are known.
\vskip 0.5cm
\begin{center}
{\bf\S 7. Stability of nonlinear shock waves}
\end{center}
\vskip 0.5cm

In Section 3, it was demonstrated that, depending on the surface
conditions, stationary nonlinear thermomagnetic shock waves of two types,
$E$ and $H$, may exist in the superconductor sample.
In this Section the stability of nonlinear shock waves
with respect to small thermal and electromagnetic perturbations in a
hard and composite superconductors is studied.
It is shown that spatially bounded solutions may correspond only to
the perturbations decaying with time, which implies stability of the
nonlinear thermomagnetic wave.

        In order to study the stability of a nonlinear wave with respect
to small perturbations, it is convenient to write a solution to Eqs. (1)-(4)
in the following form:
\beq
\begin{array}{l}
T(z,t)={T}(z)+\delta T(z,t) \exp\left[\dsf{\lambda t}{t_\kappa}\right],
\qquad\\
E(z,t)={E}(z)+\delta E(z,t) \exp\left[\dsf{\lambda t}{t_\kappa}\right].
\end{array}
\eeq
Here ${T}(z)$ and ${E}(z)$ are the stationary solutions and
$\delta T, \delta E$  are small perturbations; $\lambda$ is
a parameter to be determined.
From solution (53), one can see that the characteristic time of thermal and
electromagnetic small perturbations $t_j$ is of the order of
$t_\kappa/\lambda$.
Substituting (53) into Eqs. (1)-(4) and linearizing for small perturbations
$\left(\ds\dsf{\delta T}{T(x)}, \dsf{\delta E}{E(x)}<<1\right)$ we obtain
a differential equations for the distributions $\delta T$ and $\delta E$
in the following form
\beq
\nu\dsf{\lambda}{t_\kappa}\delta T=\dsf{\kappa}{L^2}\dsf{d^2\delta T}{dz^2}
+\left[j(z)+\sigma_{f}E(z)\right]\delta E-aE(z)\delta T\,,\\
\eeq
\beq
\dsf{1}{L^2}\dsf{d^2\delta E}{dz^2}=
\dsf{4\pi\lambda}{c^2t_\kappa}
\left[\sigma_f\delta E - a\delta T\right].
\eeq
Eliminating the variable $\delta T$ by using the relationship (55)
and substituting into Eq.(54), we obtain a
fourth-order differential equation with variable coefficients for the
distribution of small electromagnetic perturbation $\delta E$:
\beq
\dsf{d^4\delta E}{dz^4}+\beta\dsf{d^3\delta E}{dz^3}-
\left[\lambda(1+\tau)+\dsf{E(z)}{E_{\kappa}}\right]
\dsf{d^2\delta E}{dz^2}-\lambda\tau\beta\dsf{d\delta E}{dz}+
\lambda[\lambda\tau-B(z)]\delta E=0,
\eeq
Here $B(z)=\dsf{j(z)}{\sigma_f E_{\kappa}}\tau$ characterizes
the spontaneous heating of superconductor caused by a small perturbations
of $\delta T$ and $\delta E$.
The condition of existence of a non-trivial solutions of Eq. (56)
combined with boundary conditions (21) allows to define the spectrum
of eigenvalues of $\lambda$ and the stability of the thermomgnetic
wave in the sample, accordingly.
From the expression (56) just follows that if $\rm{Re}\lambda\le 0$,
then spatially bounded solutions of $\delta T$ and $\delta E$ at infinity
corresponds only a damping perturbations on the time.
This problem is complicated, and its analytical solution cannot be found in a
closed form.
Let us consider the problem of the thermomagnetic wave stability in the
limiting case of the hard $(\tau<<1)$ and the composite $(\tau>>1)$
superconductors, which is presented a greatest practical interest.
\vskip0.5cm
\begin{center}
{\bf\S 7a. Hard superconductors ($\tau<<1$)}
\end{center}
\vskip 0.5cm

As well known (see, e.g.,[8]), the magnetic flux variations in a hard
superconductors occur at a much greater rate as compared to those of the
heat transfer, so that $\tau<<1$.
The adiabatic character of the perturbations development leads to the
predominance of magnetic flux diffusion over heat diffusion in
the sample, $D_t<<D_m$.
In this case, as seen from solution (53), the characteristic
times of temperature $t_{\kappa}$ and electromagnetic field $t_m$
perturbations have to satisfy the inequalities $t_j<<t_\kappa$
$(\lambda>>1)$ and
$t_j>>t_m$ ($\lambda\tau<<1$).
Assuming that $\tau<<1$ in the limit of $\lambda>>\dsf{\nu v^2}{\kappa}$,
which corresponds to a "fast" instability [4], we obtain the equation
for determining the eigenvalues of $\lambda$
\beq
\dsf{d^2\delta E}{dz^2}+\beta\tau\dsf{d\delta E}{dz}+[B(z)-\lambda\tau]
\delta E=0,
\eeq
In hard superconductors with $\tau<<1$, the instability threshold depends
on the thermal boundary conditions at the surface of the sample only
slightly. Therefore, the electrodynamic boundary conditions at the
boundaries of the current-carrying layer $(z=+\infty, z=-\infty)$
can be neglected and one can keep only the thermal boundary conditions
to Eq. (57).
Using the substitution of variables of the type
\beq
j(z)=-\dsf{c^2}{4\pi v} \dsf{dE}{dz}= -\dsf{c^2}{4\pi v}
\dsf{\beta^2\tau E_e}{\rm{ch}^2\dsf{\beta\tau}{2}(z-z_0)},
\eeq
Eq. (57) can be rewritten in the following form [12]
\beq
\dsf{d^2\epsilon}{dz^2}+\left[\dsf{2}{\rm{ch^2}}-\Lambda\right]\epsilon=0.
\eeq
Here
$$
\Psi(z)=\delta E\exp\left[-\dsf{\beta\tau}{2}(z-z_0)\right],
$$
$$
y=\dsf{\beta\tau}{2}(z-z_0), \quad \Lambda=\lambda\tau.
$$
The Eq.(59) with the help substitution
$$
\psi(x)=(1-x^2)^{\dsf{\epsilon}{2}} \Phi(x); \quad 1-x=2s,
$$
$$
x=\rm{th}y,\quad 1-x^2=\rm{ch}^{-2} y,
$$
$$
\epsilon=\sqrt{\Lambda}, \quad P=1,
$$
can be reduced to a standard hypergeometric equation [13]
\beq
s(1-s)\dsf{d^2\Phi}{ds^2}-[\epsilon+1-2s(\epsilon+1)]\dsf{d\Phi}{ds}-
[\epsilon(\epsilon-1)-p(p-1)]\Phi=0.
\eeq
Here, the even and odd integrals can be presented as
\beq
\Phi_{1}=F(\epsilon-1, \epsilon+2, \epsilon+1, s),
\eeq
\beq
\Phi_{2}=s^{-\epsilon}F(-1, 2, 1-\epsilon, s),
\eeq
where F is the hypergeometric function [13].
Returning to the variable $y$ we have
\beq
\Psi_1(y)=(\rm{ch}y)^{-\epsilon}F\left(\epsilon-1, \epsilon+2, \epsilon+1,
\dsf{1-\rm{th}y}{2}\right)
\eeq
\beq
\Psi_2(y)=(\rm{th}y-\epsilon)\exp(\epsilon y) F\left(-1, 2, 1-\epsilon,
\dsf{1-\rm{th}y}{2}\right)
\eeq
The function $"$ must be finite
at the singular point $s=1$. The $\epsilon$ values for which $\Phi_{1}$
is finite evidently correspond to a discrete spectrum:
$\epsilon_{i}=0,1$ or $\lambda_{i}=-\tau^{-1},0.$
The function  $\Phi_{2}$ is finite only for $\epsilon=0$.
Any solution in the form of a running wave is characterized by translation
symmetry, which implies that a "perturbed"
stationary profile $E(z)$ determined by Eq. (57) corresponds to the
eigenvalue of the ground state $\lambda_{0}=0$. Differentiating Eq.
(31) with respect to $z$, one may readily see that $\dsf{dE_0}{dz}$
is an eigenfunction corresponding to $\lambda_0=0$. Indeed, the
perturbation $\delta E=\dsf{dE_0}{dz}$ essentially represents a
small wave displacement. Thus, we may suggest that tha function
$\dsf{dE_0}{dz}$ exponentially tends to zero for $z\rightarrow+\infty$
and the corresponding eigenvalue is zero. Therefore, the problem cannot
possess positive eigenvalues and $Re\lambda_{i}<0$. This result implies
that the wave is stable with respect to relatively small thermal
$\delta T$ and electromagnetic  $\delta E$ fluctuations.
The analysis of the second linearly-independent solution (64) leads to the
same conclusion.
\vskip 0.5cm
\begin{center}
{\bf \S 7b. Composite superconductors ($\tau>>1$)}
\end{center}
\vskip 0.5cm

        In composite superconductors,
an efficient value of effective electric conductivity $\sigma_f$ is much
above than in hard superconductors. As a result any intensity of the thermal
and the electromagnetic small perturbations is valid an inequality $\tau>>1$.
It may be assumed that induced the normal current $\sigma_f E$
caused by raising of temperature compensates a droping of the critical
current $j_c(T)$ and obviously, impediments of an entry of a magnetic flow
into sample.
On the other hand in composites the thermomagnetic instability development
is accompanied by "slow" perturbations with characteristic time of
a growth $t_\kappa<<t_j<<t_m$ ($\lambda<<1$) and consequently,
$t_m>>\dsf{t_m}{\lambda\tau}=\dsf{t_\kappa}{\lambda}$ ($\lambda\tau>>1$) [8].
In the approximation $\lambda<<\dsf{\nu v^2}{\kappa}$ from
Eq. (56) we obtain the following expressions for determining the
eigenvalues $\lambda$ :
\beq
\dsf{d^2\delta E}{dz^2}+\left[\beta-
\dsf{E(z)}{\beta\tau E_{\kappa}}\right]
\dsf{d\delta E}{dz}-\left[\lambda\left
(1+\dsf{E(z)}{\beta^2\tau E_{\kappa}}\right)-\dsf{1}{\tau}B(z)\right]
\delta E=0,
\eeq
The last equation with the help substitution
$$
\Psi(z)=\delta E(z) \rm{chy}(z-Z_0),
$$
can be reduced to a canonical form [12]
\beq
[H-\Lambda]\Psi=0, \\
\eeq
$$
H=\dsf{d^2}{dy^2}+U(y), \\
$$
\beq
U(y)=-1+\dsf{2}{\rm{ch^2}}+\Lambda_1 \rm{thy}; \quad y=\dsf{\beta}{2}(z-z_0),
\quad\Lambda_1=\lambda\beta^{-2}.
\eeq
The stationary wave is stable if the eigenvalues of the operator $H$
include the negative $\Lambda_1<0$. As a boundary conditions for Eq. (66)
serves the condition that the solutions $\Psi(y)$ are bounded at infinity.
The odd term in Eq. (67) describes the effect of the thermal mode on the
dynamics of electromagnetic perturbations in the superconductor.
Using the standard procedure [12] we can readily find the exact solution
of Eq. (66) in the form
\beq
\Psi(y)=(1-\rm{thy})^{p+\dsf{1}{2}}(1+\rm{thy})^{q+\dsf{1}{2}}
F\left(\alpha,\beta,\gamma,\dsf{1-\rm{thy}}{2}\right).
\eeq
Here a constant parameters $\alpha$, $\beta$, $\gamma$ are
determined by the following relationships
$$
\alpha=p+q+3,\quad
$$
$$
\beta=p+q, \quad
$$
$$
\gamma=2p+1.
$$
and
$$
p=\dsf{\sqrt{1+\Lambda_1}}{2},\quad
q=\dsf{\sqrt{1+3\Lambda_1}}{2}.
$$
It can be seen that the spectrum of eigenfunctions in (68) is
continuous [12]. Analysis of the asymptotic behaviour of solution (68)
at $y\rightarrow\pm\infty$ shows that $\delta E(y)$ is
bounded only at $\Lambda_1<0$. Indeed, one of the partial solutions
of Eq. (66), remaining bounded at $y\rightarrow+\infty$, has the
form
\beq
\delta E(y)\approx \exp(\omega_p y)\times F(p+q+2, p+q-1, 2p+1, \exp(-2y));
\quad 1-2p=\omega_p.
\eeq
It is obvious that at $y\rightarrow+\infty$
$(\exp(-2y)\rightarrow0)$ to the asymptotic (68) corresponds to
exponentially decaying at positive $\omega_{p}$ and growing wave as
$\exp(i\omega_{p}y)$ at negative $\omega_{p}$. It may be assumed that
the obtained solution exhibits regular asymptotic behaviour at
$y\rightarrow+\infty$  i.¥. the function $\delta E(y)$ is bounded only at
those eigenvalues for which $\Lambda_1<0$.
Let us represent the second linear-independent solution of Eq. (66) as:
\beq
\begin{array}{l}
F(p+q+2, p+q-1, 2p+1, s)= \dsf{ƒ(2p+1) ƒ(-2q)}{ƒ(p-q-1) ƒ(p-q+2)}\times\\
\times F(p+q+2, p+q-1, 2p+1, 1-s)+ \\
+\dsf{ƒ(2p+1) ƒ(2q)}{ƒ(p+q+2) ƒ(p+q+1)}
(1-s)^{-2q} F(p-q-1, p-q+2, -2q+1, 1-s).
\end{array}
\eeq
where ƒ is the Euler's function [13]. At $y\rightarrow -\infty$ the following
asymptotic representation is valid
\beq
\delta E(y)\approx \rm{exp}(\omega_{q}y);\quad 1-2q=\omega_q.
\eeq
It can be seen that the space-limited thermomagnetic perturbations
are damped both at $\omega_p>0$ and at $\omega_q>0$ i.e., $\Lambda_1<0$.
Let us consider a some specific cases which follows from the properties
of the asymptotic solutions of Eq. (66).\\
a)$-\dsf{1}{3}<\Lambda_1<0$; then the parameters
$\omega_p>0$ and $\omega_q>0$ can be chosen substantially.
According to the (53) small perturbations damped exponentially
with the passage of a time (see, [4]) (Fig.6).\\
b) $-\dsf{1}{3}>\Lambda_1>0$; in this specific case the wave amplitude
at the sufficiently large positive values $y$ is damped exponentially and
at the sufficiently large negative values of $y$ it has the
oscillating-damping profile with the wavelength
$L_q=\dsf{2\pi}{\rm{Jm}\cdot q}$ (Fig.7).\\
c)$\Lambda_1<-1$; in this case to the solution corresponds
oscillating-damping wave at
$y\to\pm\infty$ with the wavelength $L_q=\dsf{2\pi}{\rm{Jm}\cdot q}$ and
$L_p=\dsf{2\pi}{\rm{Jm}\cdot p}$ (Fig.8).\\

Thus it is shown that only damped perturbations correspond to
space-limited solutions, which means that the nonlinear wave is stable.
\vskip 0.5cm
\begin{center}
\bf Conclusion
\end{center}
\vskip 0.5cm

In conclusion note that the existence of essential nonlinear and
dissipative effects, connected with Joule heating at the moving of
magnetic flow, creates different regimes of automodel processes,
describing the evolution of perturbations in the vortex state of
type-II superconductor. One of measured regimes is the propagating
of thermomagnetic waves (normal region - the region heated with the
temperature, which is higher that the critical temperature $T_c$).
The nonlinear stage of evolution of thermal and electromagnetic
perturbations in superconductors is determined by the
stationary profile in the form of a running wave. It is seen that
the transition from the superconducting state to the normal state can
occur by expansion of a stationary thermomagnetic wave, which is
caused by the
balance between nonlinear effects, caused by Joule heating and dissipative
effects. The results obtained make it possible to describe the final stage
of evolution of the thermomagnetic instability in the vortex state
of type-II superconductors. It is possible that two kinds of stationary
thermomagnetic shock waves to exist, depending on the electrodynamic
boundary conditions on the sample surface.

        We should emphasis that this result was obtained
for an arbitrary temperature dependence of the thermophysical parameters
$\nu$ and $\kappa$ of superconducting material and for an arbitrary
function $j(H)$. Moreover, since the system of equations (1)-(4)
is invariant with respect to an arbitrary translation, the wave propagation
condition can be found for an arbitrary critical current density dependence
on T and H.

        Let us discuss the validity of the present approach.
The applicability of the macroscopic description is ensured by the
inequality $\Delta\xi>>d$, $d$ being the average distance between the
Abrikosov vortices. The boundary conditions used are valid to describe
the situation for a sample with a finite thickness 2L provided
$\Delta\xi<<L$. It is seen that the shock waves under consideration
can be observed only for sufficiently thick samples.

        The result obtained enables us to propose a possible version of
thermomagnetic instability development in superconductors. The initial stage
of flux jump is characterized by the exponential growth of both thermal and
electromagnetic perturbations.
The instability threshold as well
as the instability increment are defined on the basis of the linear theory
[14]. At a further stage the thermomagnetic
shock 'H-wave' with the amplitude $H_e=H_j$ spreads inside the superconductor.
Such a process results in a total penetration of the magnetic flux into
the sample. The velocity $v_H$ to be found from equation (41), determines
the time of the wave motion $t_w$ inside the sample of thickness 2L
\beq
t_w\approx \dsf{L}{v_H}.
\eeq
However, the normal transition occurring within the warmed region
complicates the condition of shock-wave propagation. A more detailed
description of a such two-phase configuration will be presented in a
further report.

        Finally, let us briefly discuss the possibility of experimental
observation of thermomagnetic shock waves. The necessary condition to observe
the phenomena predicted is to ensure sudden initial magnetic flux lattice
motion within the superconductor. For example, a thermomagnetic 'E-wave'
could be generated by initiating the magnetic flux motion in the presence
of an external magnetic field varying with the rate $\dot H_e$. Then, a
'background' electric field of order $E_b\sim \dsf{L}{c} \dot H_e$
arises inside the sample and directly determines the amplitude
of an 'E-wave' shock wave. We would point out that thermomagnetic nonlinear
waves may also be initiated by abrupt heat input leading to a sudden
temperature increase $\delta T=T^*-T_0$  at the sample surface
(accompanied by a simultaneous magnetic flux slip initiation inside
the superconductor). The electrodynamic boundary conditions
at the sample surface is determine the type ($E$ or $H$ - wave)
of the propagating wave.

\vskip 0.5cm
\begin{center}
REFERENCES
\end{center}
\vskip 0.5cm
\begin{enumerate}
\item
G.\,Nicolis, I.\,Prigozhin.
Self-organization in non equilibrium systems.
Œ.: New York, 234 (1977).
\item
A.V.Gurevich and R.G.Mints. Heat Autowaves in Normal Metals and
Superconductors. Nauka, Moscow (1987).
\item
I.L.Maksimov, Yu.N.Mastakov, and N.A.Taylanov.
Phys. Solid State,  28, 1300 (1986).
\item
N.A.Taylanov. Superconduct. Science and Technology, 14, 326 (2001).
\item
V.I. Karpman. Nonlinear Waves in Dispersive Media.
Pergamon, Oxford (1975).
\item
C.P. Bean.  Phys. Rev. Lett. 8, 6, 250 (1962).
\item
A.M.Campbell, and J.E.Evetts. Critical currents in superconductors.
London (1972).
\item
R.G. Mints, and A.L. Rakhmanov. Instabilities in Superconductors.
Nauka, Moscow, 362 (1984).
\item
P.W.Anderson, and Y.B.Kim.  Rev. Mod. Phys. 36, 39 (1964).
\item
A.A.Andronov, A.A. Vitt, and S.E.Khaykin.
Theory of oscillations. Nauka, Moscow (1981).
\item
V.I.\,Arnol'd.
Ordinary Differential Equations.
Nauka, Moscow, 264 (1971).
\item
L. D. Landau and E. M. Lifshitz.  Quantum Mechanics: Nonrelativistic
Theory (Fizmatgiz, Moscow, 1963; Pergamon, New York, 1977).
\item
D. S. Kuznetsov. Special Functions. Vysshaya Shkola, Moscow (1965).
\item
S.L.Wipf. Phys. Rev., 161, 2, 404 (1967).
\end{enumerate}
\newpage
\centerline{\large\bf FIGURE CAPTIONS}
\begin{tabbing}
Fig.1. The phase plane of Eq. (31) .\\
Fig.2. The structure of shock wave.\\
Fig.3. The phase plane of Eq. (39).\\
Fig.4. The equilibrium points of Eq. (49).\\
Fig.5. The phase plane of Eq. (49).\\
Fig.6. Asymptotical behaviour of waves at $-\dsf{1}{3}<\Lambda_1<0$.\\
Fig.7. The wave amplitude at the sufficiently large positive values\\
$y$ is damped exponentially and at the sufficiently large negative\\
values of $y$ it has the oscillating-damping profile at
$-\dsf{1}{3}>\Lambda_1>0$.\\
Fig.8. Asymptotical behaviour of waves at $\Lambda_1<-1$.\\
\end{tabbing}
\newpage
\centerline{\large \bf NIZAM A.TAYLANOV}
\begin{tabbing}
{\large \bf Address:} \\
Theoretical Physics Department and\\
Institute of Applied Physics,\\
National University of Uzbekistan,\\
Vuzgorodok, 700174, Tashkent, Uzbekistan\\
Telephone:(9-98712),461-573, 460-867.\\
fax: (9-98712) 463-262,(9-98712) 461-540,(9-9871) 144-77-28\\
e-mail: taylanov@iaph.tkt.uz \\
\end{tabbing}

\end{document}